\documentclass[aps,prd,10pt, twocolumn,twoside,floatfix,
preprintnumbers,nofootinbib]{revtex4-1}
\usepackage{amsmath, amssymb, graphics, graphicx, array, float, 
dsfont, amstext, rotating, tabularx, bbding, pifont, wasysym, 
epsfig, inputenc, bm, bbm, verbatim, color, subfigure,
dcolumn, slashed, hyperref}
\usepackage[usenames,dvipsnames]{xcolor}
\hypersetup{linkcolor=Cyan}
\newcommand{\be}{\begin{equation}}
\newcommand{\ee}{\end{equation}}
\newcommand{\bea}{\begin{eqnarray}}
\newcommand{\eea}{\end{eqnarray}}
\newcommand{\nn}{\nonumber}
\newcommand{\bi}{\begin{itemize}}
\newcommand{\ei}{\end{itemize}}
\newcommand{\bmt}{\begin{pmatrix}}
\newcommand{\emt}{\end{pmatrix}}
\newcommand{\bt}{\begin{tabular}}
\newcommand{\et}{\end{tabular}}

\newcommand{\ovl}{\overline}

\newcommand{\noi}{\noindent}
\newcommand{\ra}{\rightarrow}

\newcommand{\benu}{\begin{enumerate}}
\newcommand{\eenu}{\end{enumerate}}

\begin{document}

\title{Extended SUSY SU(5) predicting type-III seesaw testable at LHC}
\author{Ram Lal Awasthi,~Sandhya Choubey}
%\email{@hri.res.in}
%\homepage{http://www.hri.res.in}
\affiliation{\rm Harish-Chandra Research Institute, Chhatnag Road, 
Jhunsi, Allahabad 211019, India}
\date{\today}

\begin{abstract}
We propose an extension of the SUSY SU(5) which predicts LHC testable 
type-III seesaw.
The supersymmetric SU(5) GUT model 
is extended by adding a $24$-plet matter superfield along with a pair of $10_H$-plet and 
$\overline{10}_H$-plet Higgs superfields. The $24$-plet carries a triplet and a singlet fermion 
multiplet of SU(2)$_L$, which leads to type I+III seesaw. The additional $10_H$ (and 
$\overline{10}_H$) multiplets help in achieving gauge coupling unification while keeping 
the triplet fermion mass in the TeV range, making them accessible at LHC. We study the 
phenomenology of this model in detail. 
Large lepton flavor violation predicted in this model puts severe constraints 
on the Yukawa couplings of the triplet fermion. We show that this smothers 
the possibility of observing the 
contribution of the heavy fermions in neutrinoless double beta decay experiments. 
The presence of the additional $10_H$ and $\ovl{10}_H$ in this model not only 
gives gauge coupling unification, it also leads to very 
large lepton flavor violation.
 
\end{abstract}

\maketitle
\section{Introduction}

\noi
The standard model (SM) of elementary particles based on the gauge group 
SU(3)$_C \otimes$SU(2)$_L \otimes$U(1)$_Y$ is now widely accepted as 
the low energy effective theory of a more complete model of particles. 
Amongst the strongest experimental reasons demanding the extension of the 
SM are the observation of masses and mixing of neutrinos, 
presence of excess of baryons over anti-baryons in the universe, and 
the existence of dark matter and dark energy of the universe, none of which can 
be partially or wholly explained within the realm of the minimal SM. 
Amongst the theoretical reasons which beg its extension are, the 
Higgs mass and Higgs vacuum stability problem, as well as the theoretical 
prejudice that all gauge couplings will eventually unify at some high scale.  
The extensions of the SM that predict the gauge coupling unification are 
the so-called Grand Unified Theories (GUTs). The first and simplest of 
such GUT models 
proposed by Glashow and Georgi is based on the SU(5) gauge 
group \cite{Georgi:1974sy}. \\

The minimal non-SUSY SU(5)~\cite{Georgi:1974sy} model contains the 
SM fermions in three generations of 
$\ovl{5}=(\ovl{3}, 1, 1/3)\oplus(1, 2, -1/2)\equiv (d^C, L)$ and 
$10\equiv(\ovl{3}, 1, -2/3)\oplus (3, 2, 1/6)\oplus(1, 1, 1)
\equiv(u^C, Q, e^C)$, where $L$ and $Q$ are the SM SU(2)$_L$ lepton and 
quark doublets and $d^C$, $u^C$ and $e^C$ are the SM SU(2)$_L$ singlets.
The three dimensional numeric 
tuples in the above description correspond to $SU(3)_C,\,\, SU(2)_L$ 
and $U(1)_Y$ groups, respectively. 
The SM Higgs doublet is embeded in 
$5_H\equiv(3, 1, -1/3)\oplus(1, 2, 1/2) \equiv (T,H)$, where $H$ is the 
SM SU(2)$_L$ Higgs doublet while $T$ is a colored particle multiplet, absent in the SM.
The SM singlet in the adjoint representation $24_H\equiv(8,1,0)\oplus(1,3,0) 
\oplus(3,2,-5/6)\oplus(3,2,5/6)\oplus(1,1,0)\equiv(\Sigma_8, \Sigma_3, 
\Sigma_{(3,2)}, \Sigma_{(\ovl{3},2)}, \Sigma_0 )$, triggers the breaking
of SU(5) into SM, at the unification scale. The gauge bosons, as usual, are 
contained in the adjoint representation, $24_G$. While 12 of these gauge 
bosons belong to the SM ($g, ~W, ~Z$~and~$\gamma$), the rest, namely $X (Q_{EM}=4/3) 
\,\, {\rm and}\,\, Y (Q_{EM}=1/3)$, acquire masses of the order of the unification scale.
This minimal SU(5) GUT is plagued with a variety of issues. Firstly, the 
gauge couplings of the SM fail to unify at a unique point in this set-up. 
Secondly, it cannot explain non-zero neutrino masses and should 
be extended with additional multiplets to achieve it. Thirdly, there is 
nothing to remedy the gauge hierarchy problem in this set-up. It also 
fails to provide a dark matter candidate, however we will not address this 
last issue in this paper.
\\

The lacuna regarding the gauge coupling unification and the hierarchy problem 
can be easily remedied by invoking supersymmetry (SUSY) 
\cite{Dimopoulos_Georgi_1981}.  This allows to achieve the minimal supersymmetric 
standard model (MSSM) driven gauge coupling unification around $10^{16.2}$~GeV, 
and also stabilizes the SM Higgs mass. The particle content of the the SUSY-SU(5) 
is extended to include the superpartners and an additional 
$\ovl{5}_H$ representation to avoid anomaly generation and also to generate 
masses for up-type quarks. 
\\

Gauge coupling unification can also be 
achieved in extended versions of non-SUSY SU(5) with additional multiplets. 
This is particularly relevant for models which extend the particle content of SU(5) 
by adding multiplets to explain non-zero neutrino masses. The additional 
multiplets contain the heavy particle which drives the seesaw mechanism for the 
generation of the tiny neutrino masses~\cite{{Weinberg:1979sa},{Langacker:1991an},
{seesaw},{Magg:1980ut},{Foot:1988aq}}. 
The corrections to the 
running of the gauge couplings coming from the presence of these additional multiplets 
can in some cases bring about their unification without having to implement 
SUSY. 
The extensions of $SU(5)$ in this class of models that have 
recently gained popularity are the ones which predict LHC testable seesaw.
One class of models extend the fermionic content of the minimal SU(5) by adding 
the adjoint representation 
$24\equiv(\rho_8, \rho_3, \rho_{(3,2)}, \rho_{(\ovl{3},2)}, \rho_0)$ 
\cite{Senjanovic:2007zz, Dorsner2007, Georgi_Jarlskog_Ellis}. 
The heavy SM singlet and SU(2) triplet give rise to type I + type III seesaw 
which can explain the neutrino data. Demand for gauge coupling 
unification in these models predicts a 
triplet fermion within the reach of the LHC, thus opening up the 
possibility of testing seesaw at the LHC in the context of a GUT model that 
connects it to proton decay. 
The second class of extensions encompass the models which extend the 
minimal SU(5) Higgs content with an additional symmetric representation 
$15_H=(1, 3, 1)\oplus(3, 2, 1/6)\oplus(6, 1, -2/3)\equiv(\Delta_3, 
\Delta_{(3,2)}, \Delta_6$) \cite{Dorsner:2005fq, Magg:1980ut} . 
The presence of the SU(2) triplet scalar in this 
multiplet allows for the type II seesaw mechanism to generate neutrino masses 
and mixing consistent with data. Here the triplet scalar is predicted to be in the LHC testable 
range in order to be consistent with a unified coupling at the GUT scale, hence in turn 
connecting it to bounds from proton decay. 
\\

These low energy features of such extensions 
disappear in the supersymmetrized versions of these models. 
Since the MSSM drives the gauge coupling 
unification once SUSY is invoked, it leaves almost no space for additional multiplets.
Hence, addition of such multiplets significantly below the GUT scale spoils unification and 
as a result the particles in these multiplets are forced to have masses close to the 
GUT scale. The SUSY SU(5) model extended with the fermionic adjoint 
representation $24$ giving type I + type III seesaw 
was studied in \cite{perez_adj_susy_su5,Cooper:2010ik}, while 
the phenomenology of the 
SUSY SU(5) model with the symmetric $15_H$ Higgs extension giving type-II seesaw 
was discussed in 
\cite{Anna_Rossi:2002}. The seesaw scales in all these studies are 
usually at very high and hence their testability unforeseeable in any experiment. 
Also, addition of larger representations around TeV scale could lead to divergence 
of the gauge couplings before any unification. In this   
paper, we propose a model which not only predicts TeV scale type I and type III 
seesaw driven by particles that can be produced and probed at the LHC, 
but also contains a charged scalar singlet, with mass at the LHC testable scale.  
This is accomplished by adding a matter chiral superfield in the adjoint representation 
$24$ and a pair of Higgs chiral superfields in the antisymmetric $10_H$ and $\ovl{10}_H$ 
representations, to the minimal SUSY SU(5) field content. The fermions in 
$24$ lead to type I+III seesaw mechanism, while the presence of $10_H$ and 
$\ovl{10}_H$ allows us to modulate the running of the 
gauge coupling such the we get unification in the SUSY SU(5) GUT model 
with masses of the seesaw fermions in the TeV range. 
We study the phenomenology of this model in detail. 
\\

The paper is organized as follows. We begin by describing our model in Section 
\ref{sec:model}. In Section \ref{sec:gauge} we discuss gauge coupling unification
and proton decay 
and the corresponding constraints on the particle mass spectrum. In Section 
\ref{sec:seesaw} we discuss the neutrino mass generation through the type I 
+ type III seesaw mechanism. Constraints and predictions from 
lepton flavor violation and neutrinoless 
double beta decay are given in Sections \ref{sec:lfv} and \ref{sec:0vbb}, 
 respectively. Possibility of leptogenesis in this model is discussed in 
Section \ref{sec:lepto}. We end in Section \ref{sec:concl} with our conclusions.

%%%%%%%%%%%%%%%%%%%
\section{\label{sec:model} Model}
%%%%%%%%%%%
\noi
We propose an extension of the SUSY SU(5) where the field content with 
three generations of 
matter superfields $\ovl{5}$ and $10$, gauge superfield
$24_G$, and scalar superfields $5_H$, ${\ovl{5}}_H$ and $24_H$ of the minimal 
model, is augmented with one generation of matter superfields in the adjoint representation $24$, 
and a pair of scalar superfields in $10_H$ and $\ovl{10}_H$ antisymmetric representations. 
Note that our model is a further extension of the minimal adjoint SUSY SU(5) 
model proposed in the literature, with the addition of the antisymmetric 
$10_H$ and $\ovl{10}_H$. The SM multiplet within $10_H$ are,
$10_H\equiv ((1,1,1)\oplus (\ovl{3},1,-2/3)\oplus (3,2,1/6))\equiv (\chi_S,\chi_T,\chi_Z)$.
We need both $10_H$ and $\ovl{10}_H$ 
for anomaly cancellation. Neutrino masses are generated via the 
type I + type III seesaw mechanism in the same way as in the adjoint seesaw scheme 
\cite{Senjanovic:2007zz, Dorsner2007, Georgi_Jarlskog_Ellis}. 
Since we have only one $24$-plet in this model, the neutrino mass matrix 
is of rank one at the renormalizable level. However, as has been noted before, 
this problem can be cured by including higher dimensional operators, 
to increase the rank of neutrino mass matrix from one to two, allowing for 
two massive neutrino states. Higher dimensional operators are required anyway 
in SU(5) to avoid $m_d=m^T_e$, which is experimentally untenable, both in 
the non-SUSY \cite{Georgi:1974sy}
and SUSY \cite{Dimopoulos_Georgi_1981} versions of this GUT model. 
In what follows, we will see that in the model that we propose, 
the higher dimensional operators will also play a vital role in 
generating different mass scales for the additional multiplets, 
$(24, 10_H, \ovl{10}_H)$, and will be crucial for achieving the gauge coupling 
unification with the constraint that the masses of the seesaw mediating particles 
%triplet $\rho_3$ and singlet $\rho_0$ 
are within the LHC testable regime. 
And finally, the higher dimensional 
operators are also needed to create the mass separation between the SU$_L$ multiplets 
within $24$, $10_H$ and $\ovl{10}_H$, allowing for 
the triplet fermion $\rho_3$, the singlet fermion $\rho_0$ and the singly charged scalars 
$\chi_S$ and $\ovl{\chi}_S$ to be in the few 100 GeV to 1 TeV mass regime, allowing the possibility 
of producing them at the LHC. 
\\

The mass and Yukawa part of the Lagrangian involving the new field $10_H$ and $\ovl{10}_H$ 
over and above those present in the earlier versions of the minimal adjoint SUSY SU(5) are 
\bea 
{\cal L}&=& Y_{\chi}\ovl{5}10_H\ovl{5}^T+Y'_{\chi}10\ovl{10}_H24
+m_\chi{\rm Tr}(\ovl{10}_H10_H)\nn\\
&+&\lambda'_\chi{\rm Tr}(10_H\ovl{10}_H24_H)
+\lambda''_\chi{\rm Tr}(10_H24^T_H\ovl{10}_H)\nn\\
&+&\mu_{\chi}\ovl{5}_H10_H\ovl{5}^T_H+\mu_{\ovl{\chi}}{5}^T_H\ovl{10}_H{5}_H
\label{eq:lag_chi_part}
\eea 
where the first two terms on the right-hand side (RHS) are Yukawa couplings of 
the antisymmetric $10_H$ and $\ovl{10}_H$ 
superfields with the matter superfields in $\ovl{5}$, $10$ and $24$ representations. 
Rest of the terms on the RHS of the above equation are mass and Yukawa 
interactions with other scalar superfields. Terms in the last line of the equation disappear due to their antisymmetric nature. 

%%%%%%%%%%%%%%%%%%%%%%%%%%%%%%%%
\section{\label{sec:gauge} Particle mass spectrum, gauge coupling unification 
and proton decay}
%%%%%%%%%%%%

\noi
Running of the SM gauge couplings from electroweak scale up to the unification scale 
in the MSSM scenarios, following a grand SUSY desert between the TeV and GUT scales, 
gives gauge coupling unification at $\sim 10^{16.2}$~GeV. This unification 
forbids the extensions of MSSM particle spectrum at low energy scales, 
partly because the additional particles contribute to the running of the 
gauge couplings and hence spoiling unification and partly because the 
gauge couplings could start diverging even before any unification is achieved.  
Thus, the conventional SUSY SU(5) models forbid any LHC predictable seesaw scale. 
In order to alleviate this problem and make seesaw testable at LHC, we keep 
a light SU(2)$_L$ triplet fermionic superfield, $\rho_3\subset 24$.
The present bound on triplet fermion from CMS experiment is 
$m_{\rho_3}>180-210$~GeV ~\cite{CMS_EXO}. This  triplet superfield will increase the slope 
of the SU(2)$_L$ gauge coupling, thereby misaligning the 
confluence at unification scale. To compensate for that, we require an equivalent 
change in SU(3)$_C$  and U(1)$_Y$ couplings. 
In what follows, we will see that the adjustment to the SU(3)$_C$ gauge coupling is made by making 
$m_{\rho_8}= 10^{6}$~GeV, while the running of the U(1)$_Y$ gauge coupling
is tuned by demanding $m_{\chi_s+\ovl{\chi}_{{s}}}= 1$~TeV. 
\\

\begin{figure}[h!]
\includegraphics[height=7cm, width=9.0cm, angle=0]{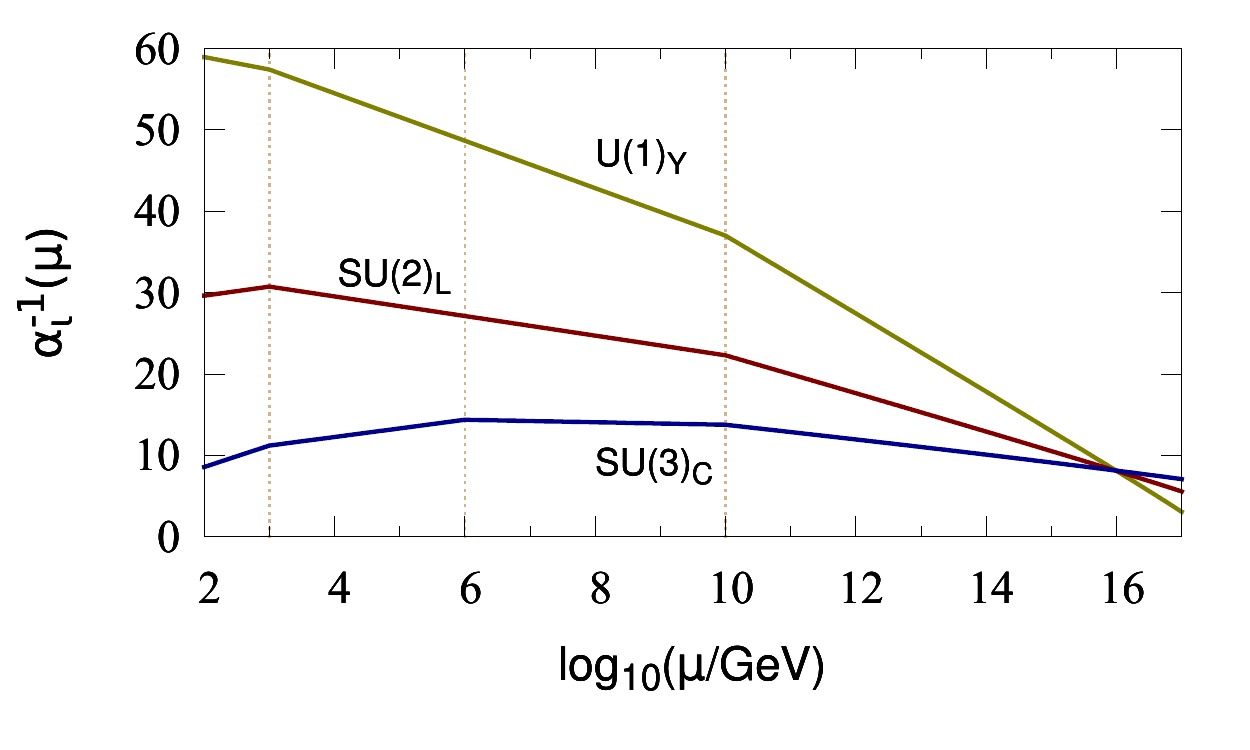} 
\caption{Two loop gauge coupling unification. The seesaw scale and 
$m_{\chi_s}=m_{\ovl{\chi}_s}$ are kept 
at TeV scale, while lepto-quarks have 
mass~$m^{lq}_\rho\simeq 10^{10}$~GeV yielding the 
unification at $M_G\simeq 10^{16}$~GeV }
\label{fig:gcu}
\end{figure}

\begin{table}[h!]
\begin{center}
\begin{tabular}{|c|c|c|c|}\hline
&&&\\[-2mm]
$\rho_3$ & $\rho_8$& $\rho_{3,2}+\rho_{\ovl{3},2}$& $\chi_S+\ovl{\chi}_S$\\[1mm]
\hline 
&&&\\[-2mm]
$\bmt 0&0&0\\ 0&24&0\\0&0&0 \emt$ & 
$\bmt 0&0&0\\ 0&0&0\\0&0&54 \emt$ &
$\bmt 25/3 & 15 & 80/3\\ 5 & 21 & 16 \\ 10/3 & 6 & 68/3 \emt$ &
$\bmt 72/5&0&0\\ 0&0&0\\0&0&0 \emt$ \\[-3mm]
& & & \\[1mm]
\hline 
\end{tabular}
\caption{\label{tab:2beta}
Two loop beta coefficients for beyond MSSM particles.}
\end{center}
\end{table}

The one loop beta coefficients for renormalization group evolution of 
gauge couplings which are active below GUT scale 
are, in SU(3)$_C\otimes$SU(2)$_L\otimes$U(1)$_Y$ format  
\bea 
&&b_{\rm SM}=(-7, -19/6, 41/10),\nn \\
&&b_{\rm MSSM}=(-3, 1, 33/5,),\nn\\
&&b_{\rho_3}=(0, 2, 0),\,\, b_{\chi^s+\ovl{\chi}^s}=(0, 0, 6/5), \nn\\
&&b_{\rho_8}=(3, 0, 0), \,\, b_{\rho_{3,2}+\rho_{\ovl{3},2}}=(2, 3, 5).
\eea 
The corresponding two loop beta coefficients for Standard Model and Minimal Supersymmetric
Standard Model can be read from~\cite{DRT_Jones} and~\cite{Martin_Vaughn:2008}, while 
two loop beta coefficients for
additional multiplets, present in the model, are listed in Table \ref{tab:2beta}. 
The two loop running of the gauge couplings from the electroweak scale to the GUT 
scale is shown in Fig.~\ref{fig:gcu}. 
The particle spectrum of our model is shown in Table~{\ref{tab:bmssm_spectra}}. 
Since 
$\rho_3$ and $\rho_8$ masses are at low energy scales, mass of the 
lepto-quarks $\rho_{(3,2)}$ and $\rho_{(\ovl{3},2)}$ is pushed at 
$m_{\rho_{(3,2)}, 
\rho_{(\ovl{3},2)}}\leq M_G^2/\Lambda$~\cite{Senjanovic:2007zz}, where $M_G$ is the unification scale 
and $\Lambda$ is the energy scale where further new physics is expected to 
take over, like Planck or string scale. 
We observe that the changes in mass scale of the 
lepto-quarks $\rho_{3,2}$ and $\rho_{(\ovl{3},2)}$ 
don't change the value of the gauge coupling, $\alpha_G$ at unification. 
The change in mass scale of lepto-quarks also doesn't alter the masses of other
fields, very significantly. However, the unification scale $M_G$ depends on the value of $m_{\rho_{(3,2)}}$ 
and $m_{\rho_{(\ovl{3},2)}}$. The masses of the lepto-quarks  
$m_{\rho_{(3,2)}, 
\rho_{(\ovl{3},2)}}=10^{10}$ GeV in our model while $m_{\rho_8}=10^6$ GeV. 
Note that the constraint $m_{\rho_8}>10^5$~GeV 
coming from cosmological bounds is 
consistent with our model~\cite{bound_on_octet}. 
\\

\begin{table}[h!]
\centering
\begin{tabular}{|r|l|}
\hline\hline &\\[-2mm]
$\rho_3\subset{24}$& $\sim 1$~TeV\nn\\[1mm]  
\hline &\\[-2mm]
$\chi_{S}+\ovl{\chi}_{S}\subset{10_H+\ovl{10}_H}$& $\sim 1$~TeV \nn\\[1mm]
\hline &\\[-2mm]
$\rho_8\subset{24}$& $\sim 10^6$~GeV\nn\\[1mm]
\hline &\\[-2mm]
$\rho_{3,2}+\rho_{\ovl{3},2}\subset{24}$& $\leq M_G/100$~GeV\\[1mm] 
\hline\hline 
\end{tabular}
\caption{BMSSM particle mass spectrum for type-III seesaw at TeV scale 
in our extended SUSY SU(5) model.}
\label{tab:bmssm_spectra}
\end{table}

\begin{figure}[t]
\includegraphics[height=7cm, width=9cm, angle=0]{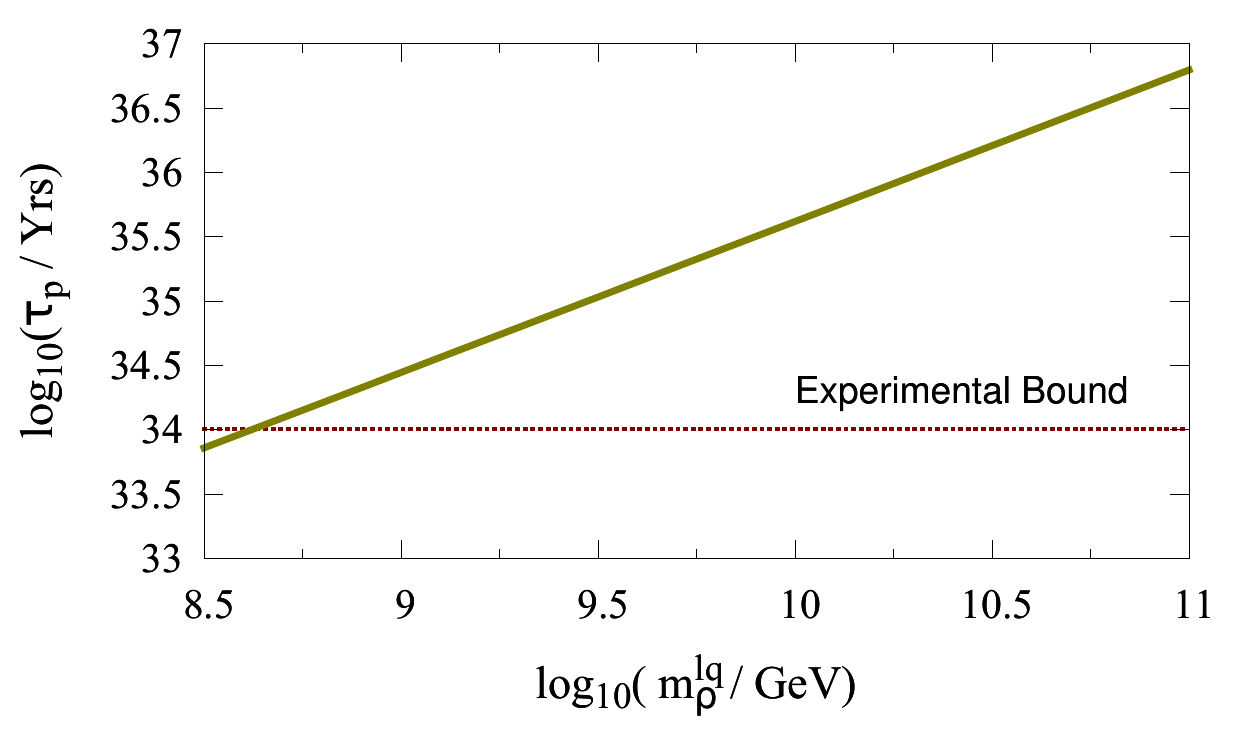} 
\caption{Proton life-time for a range of allowed lepto-quark, 
$\rho_{(3,2)}/\rho_{(\ovl{3},2)}$, masses, while GUT scale gauge coupling and 
color octet, $\rho_8$, mass remain fixed throughout the variance.}
\label{fig:life_time}
\end{figure}

From the predicted unification scale $M_{G}$, unified gauge coupling $\alpha_G$ 
at $M_{G}$ and the above mentioned particle content, we estimate the proton decay life time from the 
expression~\cite{proton_decay_weinberg}
\bea 
\Gamma^{-1}_{p\rightarrow e^+\pi^0}
&=& (1.01\times 10^{34} {\rm Yrs})\left(\frac{0.012~{\rm GeV}^3}{\alpha_H}
\right)^2\left(\frac{0.843}{A_R}\right)^2\nn\\
&\times&\left(\frac{\alpha^{-1}_G}{9.1}\right)^2\left(\frac{7.6}{F_q}
\right)\left(\frac{M_G}{3.63\times 10^{15} {\rm
GeV}}\right)^4, \label{eq:prton_decay_width} \\
&\simeq & 4.17\times 10^{35}~{\rm years}, {\rm for}\, M_G\simeq10^{16}~{\rm GeV}\, 
%{\rm inscribed\, in\, Fig.~{\ref{fig:gcu}}}
\eea
where $F_q=2(1+|V_{ud}|^2)^2\simeq 7.6$, $A_R=0.831$~\cite{anamalous_dimension} 
and $\alpha_H=0.012$~${\rm GeV}^3$~\cite{lattice_gauge},
are renormalization factor and hadronic matrix element, respectively. 
Here we assume that the mechanism suggested in  \cite{Pran_nath_dim5} is 
operative to suppress Higgsino mediated dim-5 proton decay channels.
The variation of the proton lifetime with respect to the lepto-quark 
mass scale is depicted in Fig. {\ref{fig:life_time}}.  
For $m_{\rho_{(3,2)}}=10^{10}$ GeV, the proton lifetime is predicted to be $\tau_p\sim 10^{35.6}$ years, in the range that can be probed in the future~\cite{future_proton_decay}.
On the other hand we find that the present experimental bound 
on proton life-time $\Gamma^{-1}_p> 1.01\times 10^{34}$~Yrs~\cite{superK_proton_bound}
constrains lepto-quark mass to be $m_{\rho_{3,2}}>5\times 10^{8}$ GeV. 
Hence the lepto-quarks are far beyond any direct experimental reach. 
\\

The particle spectrum of the $24$ fermionic multiplet has been discussed in detail 
in the literature before \cite{Senjanovic:2007zz}. A short mention of the 
mass spectrum of the $10_H$ multiplet is in order. From gauge coupling unification
only the ${\chi}_S$ and $\ovl{\chi}_S$
masses are required to be at low energy scale. This can be easily achieved by 
fine tuning the Eq.~(\ref{eq:lag_chi_part}) and non-renormalizable terms are 
not required. The mass spectrum for $10_H$ fields is 
\bea
m_{\chi_S}=m_\chi-6\bm{\wedge} = m_{\chi_T}-10\bm{\wedge}
= m_{\chi_Z}-5\bm{ \wedge}
\eea
where $\bm{{\wedge}}={(\lambda'+\lambda'')M_G}/{\sqrt{60}}$.

%%%%%%%%%%%%%%%%%%%%%%%%%%%%%%%%%%
\section{\label{sec:seesaw} Type-III seesaw Phenomenology and Signal at LHC}
%%%%%%%%%%%%%%%%%%%

\noindent
The complete neutrino mass matrix including singlet and neutral triplet 
fermion can emerge from the effective Lagrangian
\be 
{\cal L}_\nu=L_i(y^i_{\rho_3}\rho_3+y_{\rho_0}^i\rho_0)H 
+ \frac{1}{2}m_{\rho_3}{\rho_3}\rho_{3}
+ \frac{1}{2}m_{\rho_0}{\rho_0}\rho_0\,,
\ee 
which gives
\bea
M_\nu=\bmt 0 & m^T_D \\
m_D & M_\rho \emt\,,
\label{eq:nu_mass_matrix}
\eea
where $m_D=Y_D v_U$, $v_U$ being the vacuum expectation value of 
$H_U$ while 
\bea Y_D=
\bmt y_{\rho_0}^1 & y_{\rho_0}^2 & y_{\rho_0}^3 \\  y_{\rho_3}^1 
& y_{\rho_3}^2 & y_{\rho_3}^3 \emt,\,\, 
M_\rho = \bmt m_{\rho_0} & 0 \\ 0 & m_{\rho_3} \emt,
\eea
and light neutrino masses are written as 
\bea 
m_\nu\simeq -m^T_D M^{-1}_\rho m_D.
\label{eq:mnu}
\eea
This matrix is a complex symmetric matrix and is diagonalised by a 
unitary matrix which we denote by $U_\nu$. 
Though $m_\nu$ is a $3\times 3$ matrix its rank=2 hence, it 
give nonzero masses to only two neutrinos.  We can write~\cite{Casas_Ibarra:2001} 
\bea 
U^*_\nu \hat{m}_\nu U^\dagger_\nu &=& - m^T_D \sqrt{M^{-1}_\rho} 
\left(m^T_D\sqrt{M^{-1}_\rho}\right)^T\,, \\
%U^*_\nu \sqrt{\hat{m}_\nu} (U^*_\nu\sqrt{\hat{m}_\nu})^T& =& - m^T_D 
%\sqrt{M^{-1}_\chi} \left(m^T_D\sqrt{M^{-1}_\chi}\right)^T \\
%m^T_D&=&-iU^*_\nu \sqrt{\hat{m}_\nu}R^T\sqrt{M_\chi} \\
m_D&=&i\sqrt{M_\rho}R\sqrt{\hat{m}_\nu}U^\dagger_\nu\,.\, %\cite{Casas_Ibarra:2001}
\eea
Here $R$ is a $2\times 3$ complex matrix such that $RR^T=I_2$. Because one light 
neutrino is massless, we can express $R$ in terms of only one complex parameter for normal hierarchy~(NH) and inverted hierarchy~(IH) as \cite{Ibarra_Ross:2004}
\bea 
R&=&\bmt 0 & \cos(z) & \pm\sin(z) \\ 0 & -\sin(z) & \pm\cos(z) \emt {\rm(for~ NH)}\,, \\
R&=&\bmt \cos(z) & \pm\sin(z) & 0 \\ -\sin(z) & \pm\cos(z) & 0 \emt{\rm(for~ IH)}\,.
\eea
Hence, $R^TR=(0, I_2)$ for~NH and $R^TR=(I_2, 0)$ for~IH. Therefore we get
Dirac-Yukawa couplings like
\bea 
y^i_{\rho_0}&=&i\sqrt{m_{\rho_0}}\left(\cos(z)\sqrt{{m_\nu}_2}
{U^*_\nu}_{i2}\right. \nn\\
&&\left. \pm \sin(z)\sqrt{{m_\nu}_3} {U^*_\nu}_{i3}\right)/v_U \,,\nn\\
y^i_{\rho_3}&=&i\sqrt{m_{\rho_3}}\left(-\sin(z)\sqrt{{m_\nu}_2}
{U^*_\nu}_{i2}\right. \nn\\
&&\left. \pm \cos(z)\sqrt{{m_\nu}_3} {U^*_\nu}_{i3}\right)/v_U \,.
\eea
\begin{figure}[h!]
\begin{center}
\includegraphics[scale=.7]{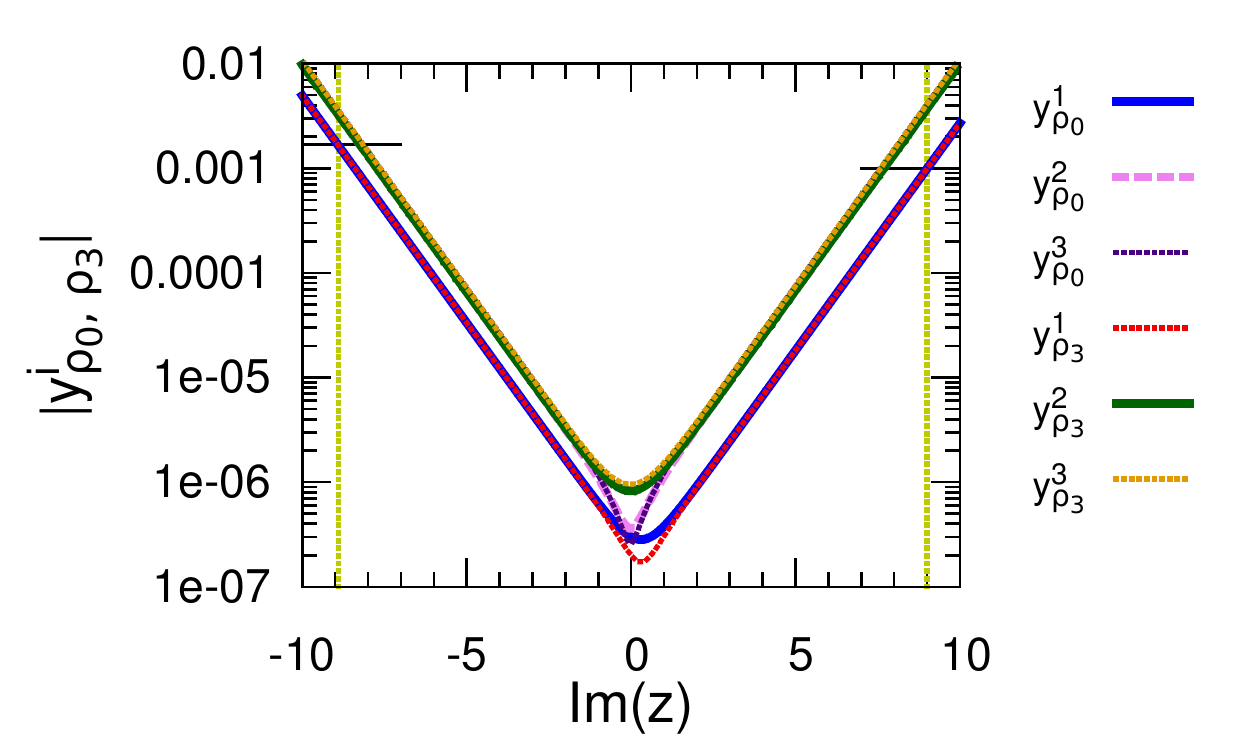}
\caption{Variation in Dirac Yukawa couplings for triplet and singlet fermions.
Horizontal line corresponds to $y_{\rho_0,\rho_3}^1=0.0017$ at $Im(z)=-9$ and 
$y_{\rho_0,\rho_3}^1=0.001$ at $Im(z)=9$. }
\label{fig:y_vs_z}
\end{center}
\end{figure}
\noi 
Similarly we can write Yukawa couplings for the inverted hierarchy (IH) case. Parameter $z$ is in general 
a complex number hence $Re(z)$ is periodic in $[0, 2\pi]$ but $Im(z)$ is free from
any constraint and gives exponential variation in $y$. The effect 
of $Re(z)$ for large $Im(z)$ becomes vanishingly small. The variation in $y^i_{\rho_3}$ 
and $y^i_{\rho_0}$ with respect to $Im(z)$ is depicted in Fig.~{\ref{fig:y_vs_z}}, where
we see that in absence of $Im(z)$, $y_{\rho_3,\rho_0}\sim {\cal O}(10^{-7}-10^{-6})$.
Together with the current status of neutrino masses and mixings~\cite{Daya_bay}, the 
only unknown parameters are the two CP-phases and a complex $z$. The measurement of 
the Yukawa couplings can constrain the $z$ parameter severely. 
\\

The triplet fermions, both charged and neutral can be produced at the LHC 
through their gauge couplings. The production rate and subsequent decay 
of these exotic fermions has been widely studied in the literature and we refer the 
reader to \cite{Senjanovic:2007zz, Del_aguila_seesaw_at_LHC,lhcpheno} for in-depth 
study of the collider phenomenology of type III seesaw at LHC. 
Here we list the decay channels of $\rho_3$, which 
depend crucially on the Yukawa couplings~\cite{Senjanovic:2007zz, Del_aguila_seesaw_at_LHC}
\bea
\label{t-z}
\Gamma(\rho_3^-\to Ze_k^-)&=&
\frac{m_{\rho_3}}{32\pi}\left|y_{\rho_3}^k\right|^2C_1C'_1,\\
\label{t0w}
\Gamma({\rho_3}^0\to W^\pm e_k^\mp)&=&
\frac{m_{\rho_3}}{32\pi}\left|y_{\rho_3}^k\right|^2 C_2C'_2,\\
\label{t0z}
\sum_k\Gamma({\rho_3}^0\to Z\nu_k)&=&
\frac{m_{\rho_3}}{32\pi}\left(\sum_k\left|y_{\rho_3}^k\right|^2\right)C_1C'_1,\\
\label{t-w}
\sum_k\Gamma({\rho_3}^-\to W^-\nu_k)&=&
\frac{m_{\rho_3}}{16\pi}\left(\sum_k\left|y_{\rho_3}^k\right|^2\right)C_2C'_2;
\eea
where $C_1=\left(1-\frac{m_Z^2}{m_{\rho_3}^2}\right)^2$,\,\, 
$C'_1=\left(1+2\frac{m_Z^2}{m_{\rho_3}^2}\right)$,\,\ 
$C_2=\left(1-\frac{m_W^2}{m_{\rho_3}^2}\right)^2$ and
$C'_2=\left(1+2\frac{m_W^2}{m_{\rho_3}^2}\right)$. This $\rho_3$ can also decay
in to Higgs and light leptons through
\bea
\label{t-h}
\Gamma\left(\rho_3^-\to h e_k^-\right)&=&\frac{m_{\rho_3}}{32\pi}\left|y_{\rho_3}^k\right|^2
\left(1-\frac{ m_h^2  }{ m_{\rho_3}^2  }\right)^2\;,\\
\label{t0h}
\sum_k\Gamma\left(\rho_3^0\to h \nu_k\right)&=&\frac{m_{\rho_3}}{32\pi}
\left(\sum_k\left|y_{\rho_3}^k\right|^2\right)
\left(1-\frac{ m_h^2  }{ m_{\rho_3}^2}\right)^2\;.
\eea
The predicted decay widths for all the above listed channels 
are given in Table~\ref{tab:decay_widths} for two cases. 
The second column of the table gives the decay width where the $Im(z)=0$ 
and we get the smallest possible value of the Yukawa couplings. Obviously this leads 
to the minimal decay width for the heavy fermion. In the third column of this table 
we give the decay widths calculated for the Yukawa couplings when $Im(z)=10$. 
The decay widths are seen to significantly increase by many orders of magnitude 
with the increase of $Im(z)$ and hence the Yukawa couplings. 
While the Yukawa couplings and the decay widths will keep increasing with the 
value of $Im(z)$, we limit our discussion to $Im(z)=10$ for reasons which will become 
clear in the following discussion on lepton flavor violation.

\begin{table}[t]
\begin{center}
\begin{tabular}{|l|c|c|}\hline
 & Decay width & Decay width\\
Decay channel & $Im(z)=0$  & at $Im(z)=10$  \\ [1mm]
$Re(z)=0$ &$\log(\Gamma_k/{\rm GeV})$ & $(x\times 10^{-4})$~GeV\\[1mm]
\hline  & & \\[-3mm]
$\sum_k{\rho_3}^-\rightarrow W^-\nu_k$ &-11 & 46.6 \\[1mm]
\hline & &\\[-3mm]
$\sum_k{\rho_3}^0\rightarrow Z\nu_k$ & -11 & 23.3\\ [1mm]
\hline & &\\[-3mm]
${\rho_3}^-\rightarrow Ze^-_k$ &(-12, -11, -11) & (0.7, 10.2, 12.4) \\[1mm]
\hline & &\\[-3mm]
${\rho_3}^0\rightarrow W^+ e_k^-$ &(-12, -11, -11) & (0.7, 10.2, 12.4)\\ [1mm]
\hline \hline & &\\ [-3mm]
${\rho_3}^-\rightarrow he_k^-$ &(-12, -11, -11) & (0.7, 9.9, 12.0)\\ [1mm]
 \hline & &\\[-3mm]
$\sum_k{\rho_3}^0\rightarrow h\nu_k$ & -11 & 22.6\\[1mm]
\hline %\hline & & \\[-3mm]
%${\rho_3}^-\rightarrow W^-{\rho_3}^0$ & &\\[1mm] \hline
\end{tabular}
\caption{\label{tab:decay_widths}Decay widths of different tree level decay channels of triplet fermions. 
The lifetime $\tau=\hbar/\Gamma$, where $\hbar=6.582119\times 10^{-23}$~GeV$\cdot$s.}
\end{center}
\end{table}

%%%%%%%%%%%%%%%%%%%%%%%%%%%
\section{\label{sec:lfv} Lepton Flavor Violation}
%%%%%%%%%%%%%%%

\begin{figure}[h!]
\begin{center}
\includegraphics[scale=.29]{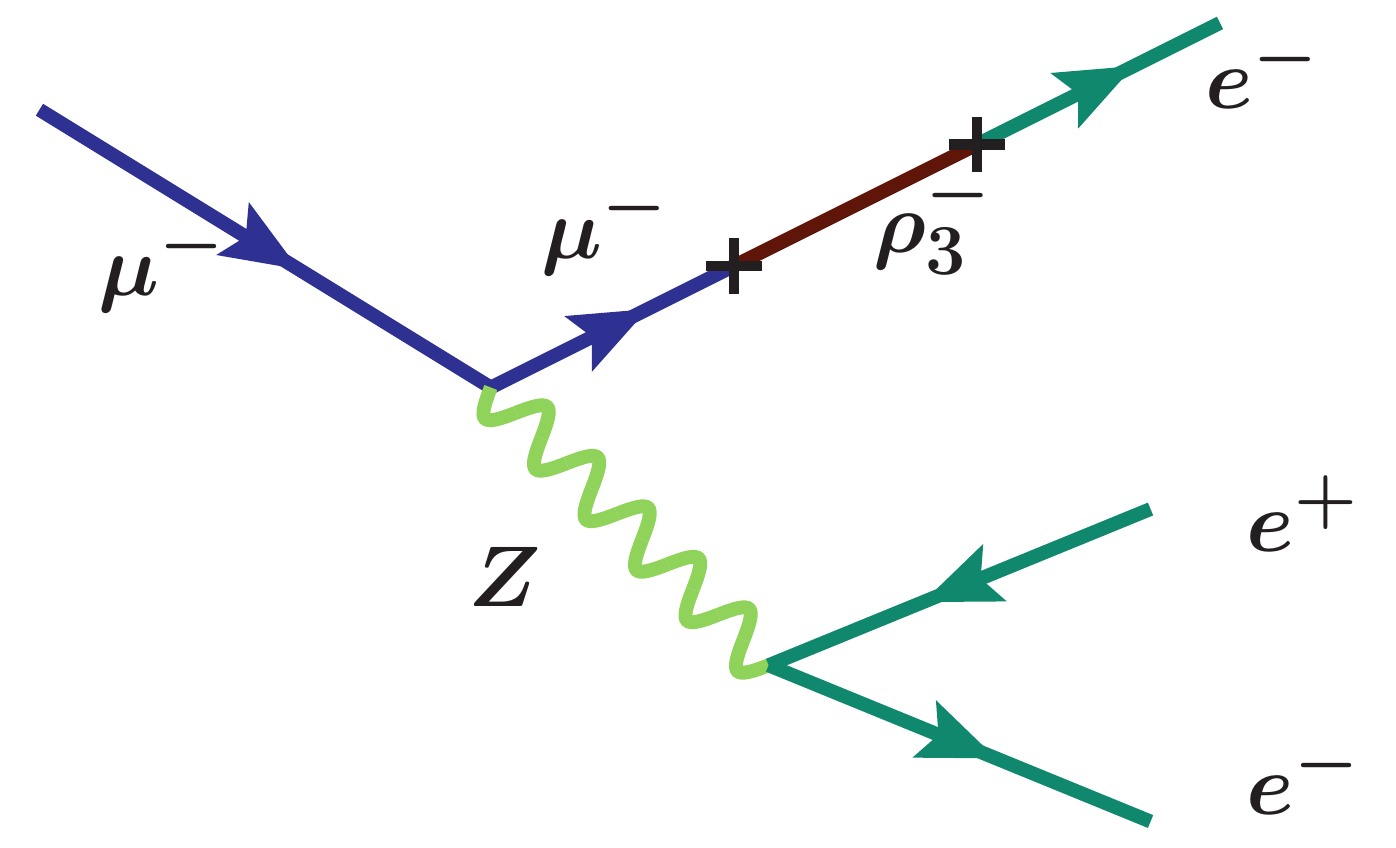}
\includegraphics[scale=.29]{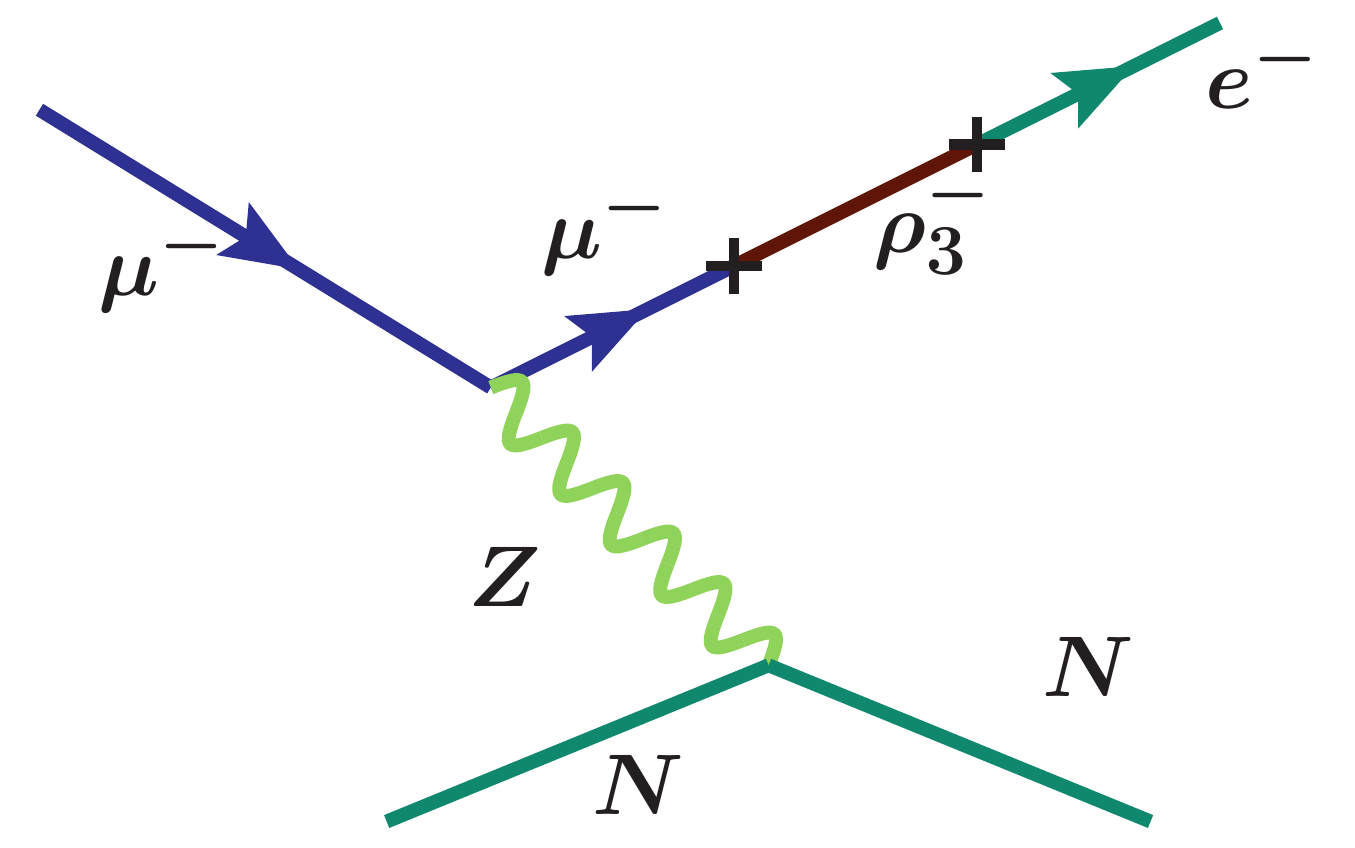}
\caption{Non-standard flavor violating decays mediated by 
the heavy charged fermion. The left panel gives $\mu \ra eee$ while 
the right panel shows $\mu\ra e$ conversion in the nucleus.}
\label{fig:lfv_mu_eee}
\end{center}
\end{figure}

\noi
Lepton flavor violation in this model is mediated even at the tree level 
by both the fermion triplet 
$\rho_3$ embedded in $24$ and the charged scalar singlet $\chi_S$ in $10_H$ of SU(5). 
The charged fermions belonging to the fermion triplet $\rho_3$ is mixed with the 
standard model charged leptons and could lead to lepton flavor changing decays 
of $\mu^-$ via the processes shown in Fig. \ref{fig:lfv_mu_eee}. The non-standard 
effective vertex shown by the crosses in the figure is lepton number 
conserving, but lepton flavor violating dimension 6 coupling, and the branching 
ratio of this process over the flavor conserving $\mu\rightarrow e\bar\nu_e\nu_\mu$ is given as~\cite{{branc_ratio_Pilaftsis},{Hambye_type3_seesaw_at_LHC}}
\bea
{\rm Br}(\mu\ra eee)=A v_U^4\left(y_{\rho_3}^\dagger\frac{1}{m^\dagger_{\rho_3}
m_{\rho_3}}y_{\rho_3}\right)^2_{\mu e}\,,
\eea
%~\cite{Hambye_type3_seesaw_at_LHC,branc_ratio_Pilaftsis}
where $A=3\sin^4\theta_W-2\sin^2\theta_W+1/2\simeq 0.1980707$. 
Imposing the current limit for the $\mu\rightarrow eee$ of \cite{mega}
\bea
{\rm Br}(\mu\ra eee)
< 1.0\times 10^{-12}\,,
\eea
gives $|y_{\rho_3}|\leq 0.0061$ for $m_{\rho_3}=1$~TeV. 
Marginally more stringent constraint comes from the experimental limits on 
$\mu \ra e$ conversion in nucleus of ${}^{48}_{22}$Ti, which gives 
\bea 
v_U^2\left(y_{\rho_3}^\dagger\frac{1}{m^\dagger_{\rho_3}m_{\rho_3}}
y_{\rho_3}\right)_{\mu e} &<& 1.7\times 10^{-7}\,.
\eea
This constrains 
$|y_{\rho_3}|\leq 0.0017$ for the same mass of $m_{\rho_3}$. These constraints 
on Yukawa couplings are stinger then constraints coming 
from ${\rm Br}(l\rightarrow l'\gamma)$.
\\

In addition to the contributions to lepton flavor violation discussed above, 
we could, in principle, have flavor violation coming from the standard 
running of the slepton mass matrices. The leading log contributions 
coming from presence of $\rho_3$ are given by~\cite{perez_adj_susy_su5}
\bea
(\delta^e_{LL})_{ij}&=&\frac{1}{8\pi^2}\frac{3m_0^3+A_0^2}{\ovl{m}^2_{\tilde{L}}}
\left(\frac{3}{2}y^{i*}_{\rho_3}y^{j}_{\rho_3}\ln\left(\frac{M_G}{m_{\rho_3}}
\right)\right. \nn\\
&&+\left.y^{i*}_{\rho_0}y^{j}_{\rho_0}\ln\left(\frac{M_G}{m_{\rho_0}}\right)\right)\,.
%(\delta^d_{RR})_{ij}&=&\frac{1}{8\pi^2}\frac{3m_0^3+A_0^2}{\ovl{m}^2_{\tilde{d}^C}}
%\left(2y^{i*}_{\rho_{3,2}}y^{j}_{\rho_{3,2}}\ln\left(\frac{M_G}{M_{\rho_{3,2}}}
%\right)\right) \,.
\eea   
Assuming $\ovl{m}^2_{\tilde{L}}\simeq m^2_0$ and $A_0\simeq 0$ we get
\be 
(\delta^e_{LL})_{ij}\leqslant {\cal O}(10^{-5})\,\,{\rm for}\,\, m_{\rho_0,\rho_3}
\simeq 1~{\rm TeV}\,.
\ee
The branching ratio of $l_i\ra l_j\gamma$ is given by \cite{Paradisi_susy_flav:2005}
\bea
\frac{Br(l_i\ra l_j\gamma)}{Br(l_i\ra l_j\nu_i\ovl{\nu}_j)}&\sim& \frac{\alpha^3}{G_F^2}
\frac{\delta^2_{ij}}{\ovl{m}^4}\tan^2{\beta} \\
&\simeq& 2.9\times 10^{-17}\,.
\label{eq:br_li_lj}
\eea
Hence we see that contribution of type III seesaw mediating $\rho_3$ to 
lepton flavor violating decays is almost negligible and can be ignored. 
\\

\begin{figure}
\begin{center}
\includegraphics[scale=.5]{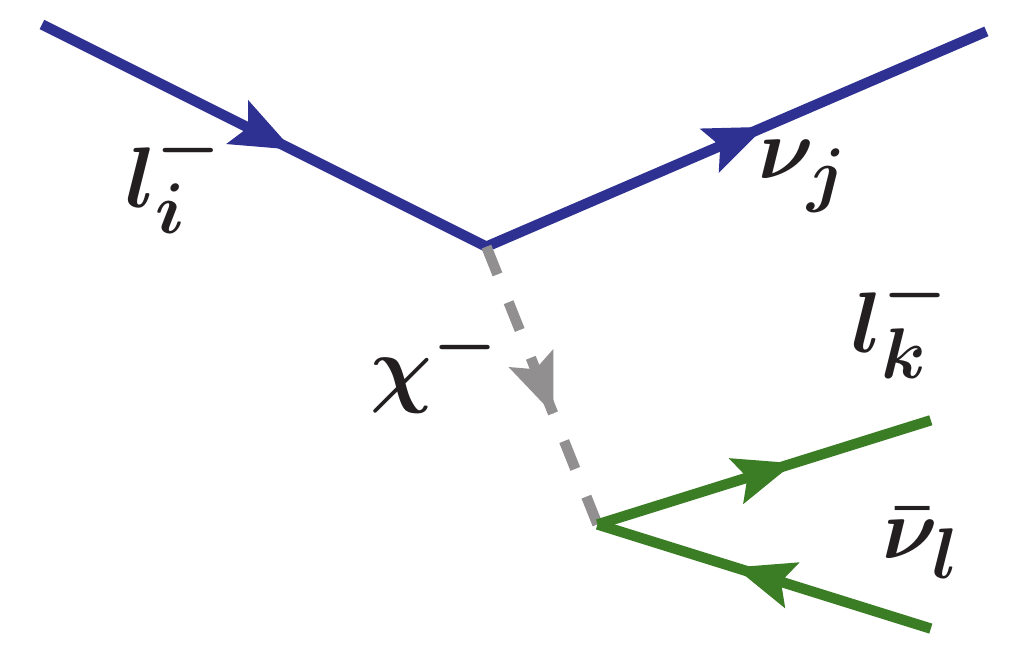}
\caption{Charged lepton decay mediated by the extra charged Higgs scalar in the theory.}
\label{fig:lepton_decay}
\end{center}
\end{figure}

Lepton flavor violation at the tree level is also induced by the singly charged 
singlet scalar $\chi_S$ in our model through diagrams shown in Fig. \ref{fig:lepton_decay}.
Note that the process $\l_i^-\rightarrow l_k^- \nu_j \bar\nu_l$ 
mediated by the charged singlet scalar $\chi_S$ comes due to the 
$l_i^-\nu_j\chi_S$ vertex from the 
first term in Lagrangian given in Eq. (\ref{eq:lag_chi_part}) which 
is a Yukawa interaction between the charged singlet scalar and the 
neutral and charged leptons of the SM. Note that this Yukawa coupling matrix 
has to be antisymmetric. Therefore, the process shown in Fig.~\ref{fig:lepton_decay} 
are necessarily flavor violating, that is, $i\neq j$ and $k\neq l$. 
Unlike the Yukawa couplings $y_{\rho_3}$ 
which are constrained by the neutrino masses, the Yukawa couplings $Y_\chi$ is 
completely unconstrained from any other process and could potentially 
lead to large flavor violating decays of the charged leptons. 
Since the two neutrinos in this diagram are not observed, this process will 
give a contribution in the experiment that observe 
the standard flavor conserving decays of the charged leptons. 
This process is similar to the one in the Zee model \cite{zee_neutrino:1980} where the 
SM is extended with an additional scalar doublet and a charged scalar singlet. 
While unlike the Zee model we do not have any radiative correction to the 
neutrino mass matrix due to the different Higgs structure and SUSY nature of 
our model, we do have the $\chi_S$ mediated 
process shown in Fig. \ref{fig:lepton_decay} as a 
result of the $Y_\chi$ Yukawa coupling in our model. 
Corrections to the four-Fermi interaction responsible for the 
weak decays such as $\mu^-\rightarrow e^-\bar\nu_e\nu_\mu$ has been  
calculated in the literature as $(G_F/\sqrt{2})(1+\zeta)$ \cite{smirnov_zee_model:1997}, 
where $\zeta$ is given as 
\be 
\zeta =\frac{a^2/2m^2_{\chi_S}}{G_F/\sqrt{2}}\,,
\ee
where $a$ is $e\mu$ element of the antisymmetric Yukawa matrix, $Y_\chi$,
which is parametrized in terms of three complex parameters as
\be 
Y_\chi=\bmt 0&a&b\\ -a&0&c\\ -b&-c&0  \emt\,,
\label{eq:antisymm_matrix} 
\ee 
and we have taken these parameters to be real for simplicity.
The experimental observations return the constraint  $\zeta<10^{-3}$, which 
leads to the corresponding constraint on the $\mu e$ element of the Yukawa matrix $Y_\chi$ as
$a^2<1.65\times 10^{-8}m^2_{\chi_S}/({\rm GeV^2})$.
For a TeV scale $m_\chi$ we get $a< 0.128$ which is large enough to 
generate significant flavor violation. The branching ratio for 
$\mu\ra e\gamma$ driven by $Y_\chi$ 
is~\cite{smirnov_zee_model:1997, meg_coll_mu_to_e_gamma}
\be 
Br(\mu\ra e\gamma)=\left(\frac{\alpha}{48\pi}\right)
\left(\frac{bc}{m^2_{\chi_S} G_F}\right)^2<5.7\times 10^{-13},
\ee  
which constrains the Yukawa couplings to $bc<0.00126$. 
Similar constraints will come from $Br(\tau\ra \mu\gamma)<4.4\times 10^{-8}$ 
\cite{tau_to_light_gamma} and $Br(\tau\ra e\gamma)<3.3\times 10^{-8}$ 
\cite{tau_to_light_gamma}, and, are $ab<0.124$ and $ac<0.093$ 
respectively. If we saturate the $\tau\ra \mu\gamma$ and $\tau\ra e\gamma $ limits, 
we have $b < 1$ and $c < 0.72$, respectively, since $a<0.128$. Therefore, 
from the bounds from radiative $\tau$ decays we have $bc<0.72$, which is 
much weaker than the bound we have on this combination from $\mu\ra e\gamma$. 
\\

In addition, we could have lepton flavor violation coming from 
the contributions  of $\chi_S$ to the running of the slepton masses which could 
lead to radiative decays of the charged leptons. The leading log contribution 
in this case is given as~\cite{Anna_Rossi:2002}
\be 
(\delta^e_{LL})_{ij}\simeq \frac{1}{2\pi^2}\frac{3m_0^2+A_0^2}{\ovl{m}^2_{\tilde{L}}}
(Y_{\chi}Y^{\dagger}_{\chi})_{ij}\ln\left(\frac{M_G}{m_{\chi_S}}\right).
\label{eq:delta_LL}
\ee 
Without any further assumption and saturating the experimental bound 
we can find that
\bea
\frac{b}{a}&\simeq& \frac{Br(\mu\ra e\gamma)}{Br(\mu\ra e\nu_\mu\ovl{\nu}_e)}\frac{Br(\tau\ra e\nu_\tau\ovl{\nu}_e)}{Br(\tau\ra e\gamma)}\,,\nn\\
\frac{c}{b}&\simeq& \frac{Br(\tau\ra e\gamma)}{Br(\tau\ra e\nu_\tau\ovl{\nu}_e)}\frac{Br(\tau\ra \mu\nu_\tau\ovl{\nu}_\mu)}{Br(\tau\ra \mu\gamma)}\,.
\label{eq:ratio_branching_ratio}
\eea
\\

The conclusion we 
can make from this analysis is that even though we do not get significant 
contribution to LFV from type III seesaw for possible parameter space, we still
may explain it due to presence of extra charged singlet scalar. Also, the 
observation of $\mu\ra e\gamma$ in near future constrains $\tau\ra \mu\gamma$
and $\tau\ra e\gamma$ to remain unobserved for very long.

%%%%%%%%%%%%%%%%%%%%%%%%%
\section{\label{sec:0vbb} Neutrinoless double beta decay}
%%%%%%%%%%%%%%

\begin{figure}[b]
\begin{center}
\includegraphics[scale=.25]{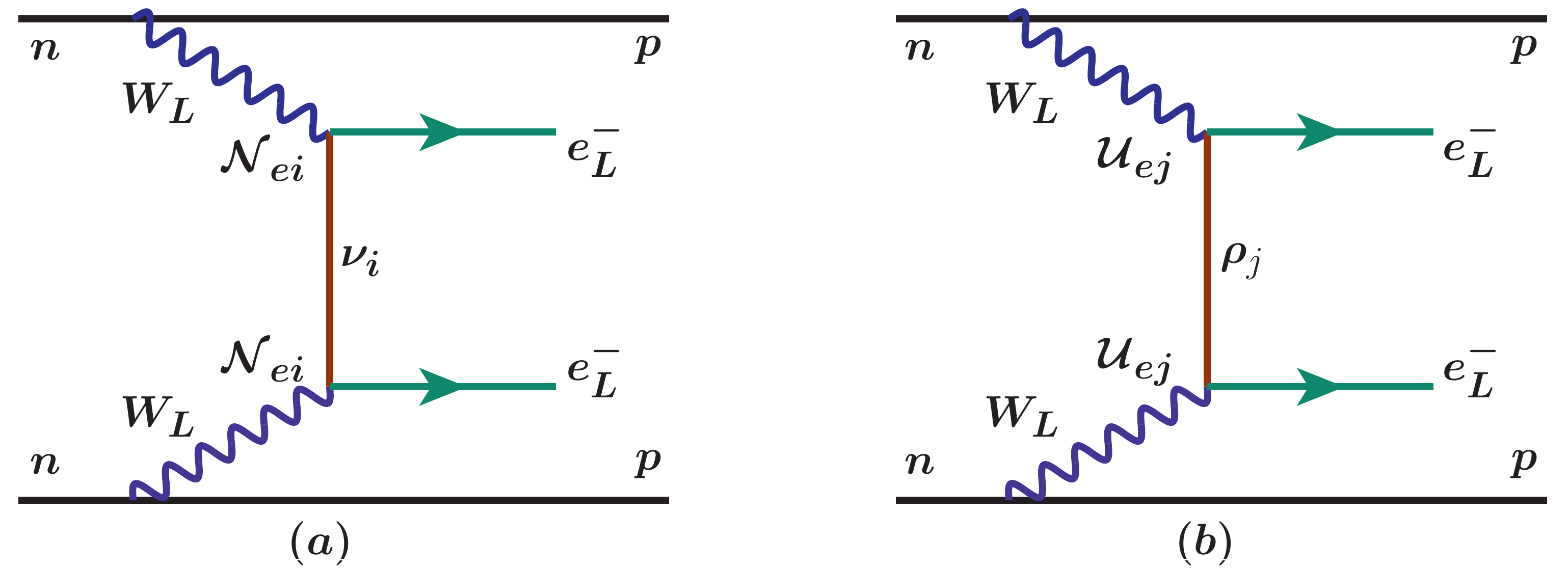}
\caption{Feynman diagrams for neutrinoless double beta decay.}
\label{fig:onbb_feyn}
\end{center}
\end{figure}

\begin{figure}[t]
\begin{center}
\includegraphics[scale=.63]{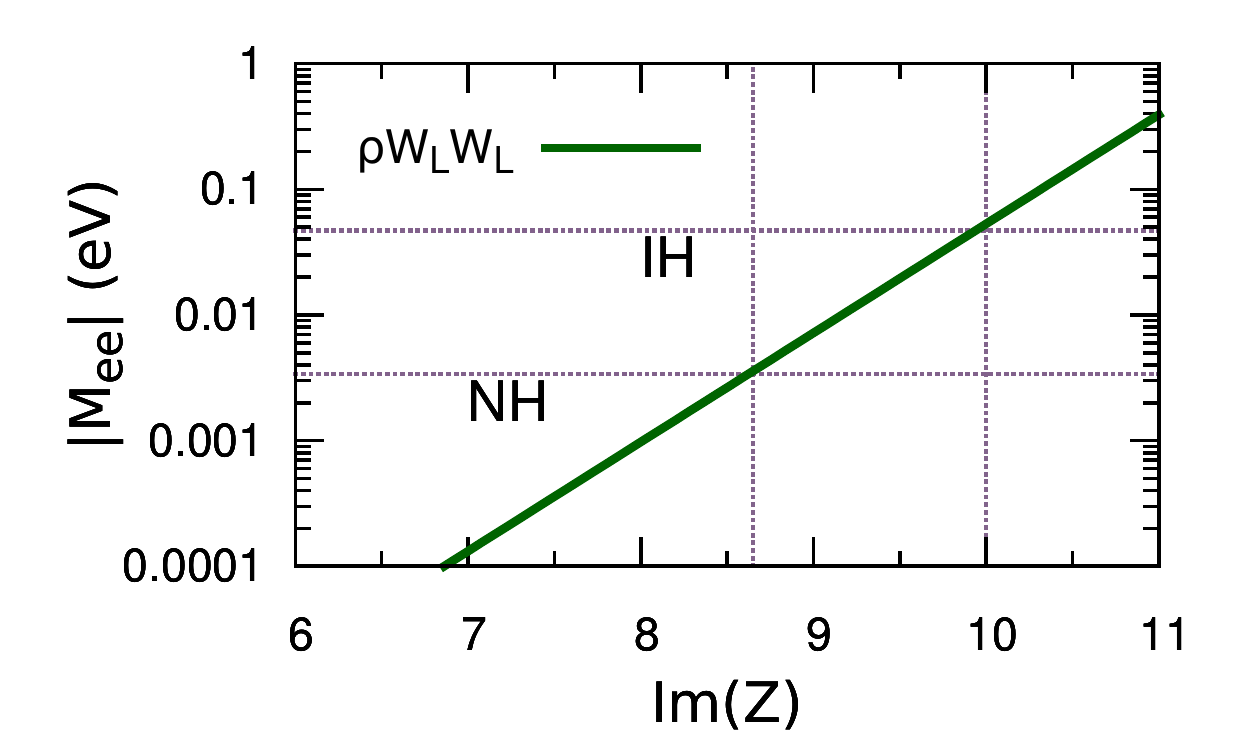}
\caption{Effective mass in $0\nu\beta\beta$-decay as a function of 
imaginary value of the seesaw parameter z. The green solid line shows the 
contribution due to the heavy states while the horizontal dotted lines 
give the contribution from the light neutrino states for normal (NH) and 
inverted (IH) neutrino mass hierarchies.}
\label{fig:onbb}
\end{center}
\end{figure}

\noi
The symmetric mass matrix $M_\nu$, in Eq.~({\ref{eq:nu_mass_matrix}}) is diagonalized by the unitary $5\times 5$ matrix~\cite{grimus_seesaw}
\bea 
W&=&\bmt {\cal N} & {\cal U} \\ {\cal T} & {\cal V} \emt \nn\\
&\simeq& \bmt 1-\frac{1}{2}BB^\dagger & B \\ -B^\dagger 
& 1-\frac{1}{2}B^\dagger B \emt
\bmt 
U_\nu &0 \\
0 & U_\rho
\emt
\label{eq:ref}
\eea
such that $W^T M_\nu W={\rm diag}(m_1, m_2, m_3, m_{\rho_0}, m_{\rho_3})$, and
$B^\dagger\simeq M_\rho^{-1}m_D$. Here $U_\nu$ diagonalizes the light 
neutrino mass matrix $m_\nu$ (cf. Eq. (\ref{eq:mnu})), while $U_\rho$ diagonalizes 
heavy neutrino mass matrix. 
\\

The contribution to
neutrinoless double beta decay~\cite{rodejohann_onbb2012} will come from the diagrams 
depicted in Fig.~{\ref{fig:onbb_feyn}}, wherein we have contributions 
from the exchange of both the light (left panel) and heavy (right panel) neutral 
fermion. The total amplitude for the process 
can therefore be written as~\cite{joydeep_chakrabortty}
\be 
{\cal A}\simeq G_F^2 \left(\sum_i\frac{{\cal N}^2_{ei}m_i}{p^2} 
-\sum_j\frac{{\cal U}^2_{ej}}{m_{\rho_j}}\right),
\ee
where $i$ and $j$ run over first three and last two indices of the above 
eigenvalues, respectively. The half life of neutrinoless double beta decay
can be written as~\cite{joydeep_chakrabortty}
\be 
\Gamma_{0\nu\beta\beta}= G \frac{|{\cal M_\nu}|^2}{m_e^2}\left|
\sum_i{\cal N}^2_{ei}m_i+p^2\sum_j \frac{{\cal U}^2_{ej}}{m_{\rho_j}}\right|^2{\ln}(2),
\ee
where $G$ contains the phase space factors, $m_e$ and ${\cal M_{\nu}}$ are  
electron mass and nuclear matrix element, respectively, and 
where $|p^2|\sim (190 ~{\rm MeV})^2$\cite{momentum_exchange} is the momentum 
exchanged in the process.
The mixing matrices ${\cal N}$ and ${\cal U}$ are given in Eq. (\ref{eq:ref}). 
Elements inside the
big modulus are the effective mass contributions to $0\nu\beta\beta$-decay
due to light neutrinos and heavy neutral fermions, respectively. In the 
Fig.~\ref{fig:onbb} we have plotted the contribution of the heavy fermions 
to the $0\nu\beta\beta$-decay effective mass, as a 
function of the seesaw parameter, $Im(z)$. 
The two horizontal lines in the figure represent the maximum achievable effective 
mass contribution due to light neutrino mediation, in normal and inverted 
hierarchy scenario, respectively. 
%The EXO experiment is proposed to explore the region in between and will lead 
%to a beyond Standard Model signature in normal hierarchy scenario of neutrino masses. 
We reiterate that the lightest of the light neutrino in our model is massless. 
We note from the figure that the contribution of the heavy 
neutrinos to the effective mass in $0\nu\beta\beta$-decay dominates the one coming 
from the light neutrinos when $Im(z) > 8.5$ and 10, for normal and inverted 
neutrino mass hierarchy, respectively. Fig. \ref{fig:y_vs_z} reveals that for 
these values of $Im(z)$, $y_{\rho_3} > 0.001$ and 0.01 
for normal and inverted neutrino mass hierarchy, respectively. However, we had seen in 
the last section that such large values of the Yukawa couplings are 
ruled out by the current constraints on lepton flavor violation. Therefore, the 
$0\nu\beta\beta$-decay signal in this model is expected to be driven by the 
standard light neutrino mass term.

\section{\label{sec:lepto}LEPTOGENESIS}

\noi
A simple explanation to baryonic asymmetry of universe is provided by the
Baryogenesis via Leptogenesis. The leptonic asymmetry is generated by out
of equilibrium decay of seesaw generating particles. Usually in type-I 
or type-III seesaw with more then one generations, TeV scale leptogenesis 
may occur through resonance in self-energy contribution~\cite{Pilaftsis_resonant}. 
But, in a hybrid seesaw through type-I+III~\cite{blanchet_perez_adj_lept:2008}, 
as we discussed above, $CP$-asymmetry is generated 
only through vertex correction, because with the present experimental bound 
on $m_{\rho_3}$ self-energy contributions are absent to avoid the braking of
$SU(2)_L$ symmetry at this scale. The vertex correction, gives asymmetry of 
the type \cite{blanchet_perez_adj_lept:2008, Hambye:2003rt}
\be 
(\epsilon_{\rho_3})_i=\frac{{\rm Im}\left[y^{*i}_{\rho_3}
y^{i}_{\rho_0}(y^\dagger_{\rho_3}y_{\rho_0})
\right]}{8\pi (y^\dagger_{\rho_3}y_{\rho_3})_{11}}
f\left(\frac{m^2_{\rho_0}}{m^2_{\rho_3}}\right)
\ee     
where %$m_{\rho_3}<m_{\rho_0}$, and
\be 
f(x)=\sqrt{x}\left[(1+x)\, {\rm ln}\left(1+\frac{1}{x}\right)-1 \right]
\ee
Substituting $\rho_3\leftrightarrow \rho_0$ will give asymmetry generated by singlet
fermion. Because $f(x)<1,\,\, \forall x$; this asymmetry does not lead to a resonance effect, 
like we get for the self-energy term. As we see in the Fig.~\ref{fig:y_vs_z} that for
$y_{\rho_3,\rho_0}>>10^{-6}$ the imaginary part is the major contribution 
to the Yukawa couplings, and that $y^i_{\rho_3}- y^i_{\rho_0}\simeq\Delta y^i$. For the Yukawa
couplings saturating the flavor violating bounds, for $m_{\rho_0}\simeq m_{\rho_3}$ we get 
\be 
(\epsilon_{\rho_3})_i \sim\frac{{\rm Re}(\Delta y^i){\rm Im}(y^i )}{8\pi}
f\left(\frac{m^2_{\rho_0}}{m^2_{\rho_3}}\right)
<< 10^{-10}.
\ee
The total asymmetry $\epsilon_{\rho_3}=\sum_i (\epsilon_{\rho_3})_i$
is constrained as ~\cite{david_ibarra_bound:2002} as
\be 
\epsilon_{\rho_3}\lesssim \frac{3}{8\pi}\frac{m_{\rho_3}m_{\nu_3}}{v^2_u}
\sim 10^{-13}.
\ee
With this small CP asymmetry we do not get any significant amount of baryonic
asymmetry, because
\bea 
\eta_B \simeq && 10^ {-2} \sum_{i}\epsilon_i \kappa_i(w\rightarrow \infty),\nn \\
{\rm while}\hspace*{1cm}\eta^{\rm CMB}_B =  &&(6.2\pm 0.15)\times 10^{-10}
\eea 
and as the efficiency factor $\kappa_i<<1$, the model prediction never 
reaches the experimental value $\eta^{\rm CMB}_B$\cite{WMAP_collab}.
Here, $10^{-2}$ is the dilution factor coming from converting the CP asymmetry to
the baryonic asymmetry and $w=m_{\rho_3}/T$, where $T$ is the temperature.
Hence, leptogenesis is expected to fail in this model and 
we need to look alternative ways to explain the baryonic asymmetry of the universe.
A significant CP asymmetry requires both real and imaginary parts of
the Yukawa couplings to be large, which is possible only when the masses of
the triplet and singlet fermion are increased, well beyond the reach of the LHC. 
Therefore, any observation of triplet 
fermion and SUSY in near future at the LHC/ILC, could be taken as a hint 
for non possibility of leptogenesis within these classes of type-I+III models.

%where
%\bea 
%\kappa_i &\simeq& \int^{w}_{w_0}dw'\frac{dN^{eq}_{\rho_i}}{dw'}\nn\\
%&\times& exp\left[-\int_{w'}^{w} dw''\left(\sum_i W_{ID}(K_i, w')\right)\delta_i^2 \right]
%\eea

%where
%\bea 
%\kappa_i &\simeq& \int^{w}_{w_0}dw'\frac{dN^{eq}_{\rho_i}}{dw'}\nn\\
%&\times& exp\left[-\int_{w'}^{w} dw''\left(\sum_i W_{ID}(K_i, w')\right)\delta_i^2 \right]
%\eea

\section{\label{sec:concl}Discussion and Conclusion}
\noi
We proposed an extension of the 
SUSY SU(5) GUT model which allows to test 
seesaw at the LHC. This is accomplished by adding 
a fermionic $24$ matter superfield and a pair of $10_H$ and $\ovl{10}_H$ 
Higgs superfield to the model. The triplet and singlet fermions from the $24$ 
representation lead to type I+III seesaw in this model. The presence of the  
charged singlet scalar in $10_H$ and $\ovl{10}_H$ allows the controlling of the 
gauge coupling running in such a way that unification is achieved in this model 
with the triplet fermions and the charged singlet scalar masses being 
around 1 TeV. This opens up the possibility of producing and testing these 
particles at the colliders. 
\\

We studied the seesaw phenomenology of this model and showed how the 
additional freedom from the seesaw framework allows for very large Yukawa 
couplings $y_{\rho_3}$ for the heavy fermions even for TeV scale masses. We parametrized this 
freedom in terms of a complex variable $z$ and showed that the $Im(z)$ 
component could give very large Yukawa couplings, leading to enhanced decay 
width and hence shorter lifetimes for the heavy fermions at the collider. We studied 
the lepton flavor violation predicted in this model. Constraints from $\mu\ra eee$ and 
$\mu \ra e$ conversion was shown to 
severely constrain the Yukawa couplings $y_{\rho_3}$ in this model. 
We also 
studied the contribution of the heavy fermion exchange in neutrinoless double beta decay 
fermions. We showed that the constraints from lepton flavor violation completely 
smother any chances of seeing these contribution in the neutrinoless double beta decay 
experiments. 
\\
 
In addition to the exotic fermion, we also have an exotic single charged scalar in this model. 
We showed that very large lepton flavor violation at the tree level is induced in this model 
by the charged singlet scalar. We studied these processes and put constraints on the 
coupling of the charged singlet scalar with the SM fermions. 
\\

In conclusion, our model predicts a testable seesaw within a SUSY GUT framework. 
In addition, it has a rich phenomenology in lepton flavor violation experiments.   
In case SUSY as well as fermionic triplets
are observed at the LHC or the future colliders, it will be a test for type-III seesaw within a SUSY framework and 
this model will be a viable candidate as the beyond the standard model theory to explain this result.

\section*{APPENDIX}
%\vspace*{1.3cm}
\subsection*{Renormalization group equations}
\noi
The one-loop beta coefficients for Yukawa couplings, soft scalar masses and trilinear
terms are standard and straightforward to 
evaluate~\cite{Martin_Vaughn:2008, Yamada_supergraph} by hand or using a 
package~\cite{Staub_Fonseca}. A sample of slepton soft mass RGE is
%\vspace*{-.8cm}
\bea
\beta_{\tilde{m}^2_L}&=&2\sqrt{3}h_{\rho_3}h^{\dagger}_{\rho_3}
+2h_{\rho_0}h^{\dagger}_{\rho_0}
+8h_{\chi} h^\dagger_{\chi}+2h_{l}h^{\dagger}_{l} \nn\\
&+& 6{\bm I}\left(\frac{1}{5}|m_{1}|^2g_{1}^{2}
-|m_{2}|^2g_{2}^{2}\right)\nn\\
&-&\frac{3}{5}g_{1}^{2}{\bm I}\left(m_{H_u}^{2}-m_{H_d}^2
-m_{\chi}^2+m_{\chi}^2
+Tr(m_{Q}^2)\right. \nn\\
&-&2Tr(m_u^2)-\left. Tr(m_d^2)-Tr(m_L^2)+Tr(m_e^2)\right)\nn\\
&+&\sqrt{3}\left(2m_{\rho_3}^2y_{\rho_3}y^\dagger_{\rho_3}
+y_{\rho_3}y^\dagger_{\rho_3}{m^2_{L}}
+{m^2_{L}}y_{\rho_3}y^\dagger_{\rho_3}\right)\nn\\
&+&2m_{H_u}^2y_{\rho_0}y^\dagger_{\rho_0}+ 2m_{\rho_0}^2y_{\rho_0}y^\dagger_{\rho_0}
+y_{\rho_0}y^{\dagger}_{\rho_0}{m_{L}^2}\nn \\
&+&{m_{L}^2}y_{\rho_0}y^{\dagger}_{\rho_0}+8m_{\chi_S}^2Y_{\chi}Y^{\dagger}_{\chi}
+4Y_{\chi}Y^{\dagger}_{\chi}{m^2_L}\nn\\
&+&8Y_{\chi}{{m^2_L}^T}Y^\dagger_{\chi}+4{m^2_L}Y_{\chi}Y_{\chi}^{\dagger}
+2m^2_{H_d}y_{l}y^{\dagger}_{l}\nn\\
&+&y_ly_l^{\dagger}{m^2_{L}}+2y_l{m^2_{e}}^Ty_l^{\dagger}+{m^2_{L}}y_ly_l^{\dagger}.
\eea
Experimental bound on LFV processes constraints the couplings of triplet and singlet fermionic superfields to be very small. Hence, we explicitly write the 
new contributions to soft mass beta coefficients due to $\chi_S, \ovl{\chi}_S$ only. 
With the constraint $y_{\rho_3, \rho_0}<<Y_{\chi}$, the leading new contributions 
beyond MSSM are
\bea
\beta^{\chi}_{m^2_Q}&=&0,\nn\\
\beta^{\chi}_{m^2_u}&=&\frac{4}{5}g^2_1{\bm 1}m^2_{\bar{\chi}}
-\frac{4}{5}g^2_1{\bm 1}m^2_{\chi},\nn\\
\beta^{\chi}_{m^2_d}&=&-\frac{2}{5}g^2_1{\bm 1}m^2_{\bar{\chi}}
+\frac{2}{5}g^2_1{\bm 1}m^2_{\chi},\nn\\
\beta^{\chi}_{m^2_e}&=&2h'_{\bar{\chi}}{h'}^{\dagger}_{\bar{\chi}}
-\frac{6}{5}g^2_1{\bm 1}m^2_{\bar{\chi}}+\frac{6}{5}g^2_1{\bm 1}m^2_{\chi}
+2m^2_{\bar{\chi}}Y'_{\bar{\chi}}{Y'}^{\dagger}_{\bar{\chi}}\nn\\
&+&2m^2_{\rho_0}Y'_{\bar{\chi}}{Y'}^{\dagger}_{\bar{\chi}}
+2Y'_{\bar{\chi}}{Y'}^{\dagger}_{\bar{\chi}}m^2_{e}
+2m^2_{e}Y'_{\bar{\chi}}{Y'}^{\dagger}_{\bar{\chi}}\nn\\
\beta^{\chi}_{\tilde{m}^2_L}&\simeq&8h_{\chi} h^\dagger_{\chi} 
+8m_{\chi_S}^2Y_{\chi}Y^{\dagger}_{\chi}
+4Y_{\chi}Y^{\dagger}_{\chi}{m^2_L}\nn\\
&+&8Y_{\chi}{{m^2_L}^T}Y^\dagger_{\chi}+4{m^2_L}Y_{\chi}Y_{\chi}^{\dagger}
\label{eq:beta_lfv_cont}
\eea

\vskip 1cm
%\newpage
\noindent
%%%%%%%%%%%%%%%%%%%%%%%%%%%%%%%%%%%%%%%%%%%%
{\large \bf Acknowledgements}
%%%%%%%%%%%%%%%%%%%%%%%%%
\\

S.C. acknowledges partial support from the European Union FP7 ITN INVISIBLES
(Marie Curie Actions, PITN-GA-2011-289442).

\end{document}